\documentclass{aa}
\usepackage{graphicx}
\begin{document}

   \title{Detection of a classical $\delta$ Scuti star in the new eclipsing binary system 
    HIP~7666
}
   \author{E. Escol\`{a}-Sirisi \inst{1}
          \and J. Juan-Sams\'{o} \inst{1}
          \and J. Vidal-S\'{a}inz \inst{1}
          \and P. Lampens \inst{2}
          \and E. Garc\'{\i}a-Melendo \inst{3}
          \and J. M. G\'omez-Forrellad \inst{1,3}
          \and P. Wils \inst{4}
          }

   \offprints{E. Garc\'{\i}a-Melendo}

   \institute{Grup d'Estudis Astron\`{o}mics, Apdo. 9481, 08080 Barcelona, Spain\\
             \email{duranobs@astrogea.org}
             \and
             Koninklijke Sterrenwacht van Belgi\"e,
             Ringlaan 3, B-1180 Brussel, Belgium\\
             \email{patricia.lampens@oma.be}
             \and
             Esteve Duran Observatory Foundation, Montseny 46 -- Urb.\ El Montanya, 08553 Seva, Spain\\
             \email{duranobs@astrogea.org}
             \and
             Vereniging Voor Sterrenkunde, Belgium\\
             \email{patrick.wils@cronos.be}
             }

   \date{Received , 2003; accepted , }

   \abstract{
   HIP 7666 is a variable star newly discovered during the Hipparcos mission and classified as of unknown type (ESA \cite{esa97}).
During 23 nights between July 2000 and November 2000, over 2300 CCD observations in the V band were obtained. These data show that the new variable is a detached eclipsing binary system with an orbital period 
of 2.37229 days. In addition, one of the components undergoes very short-period oscillations with a main pulsation frequency of 
24.46 or 25.47 c/d. HIP~7666 is therefore a new member of the rare group of detached eclipsing binary systems with 
a $\delta$ Scuti type component.
   }

   \maketitle
   
   \authorrunning{Escol\`{a}-Sirisi, E. et al.}
   \titlerunning{A $\delta$ Scuti star in the new eclipsing binary system HIP 7666}

\section{Introduction}
HIP 7666 (=~HD~232486~=~GSC~3671-1094~=~NSV~15349) is a new variable star discovered by the Hipparcos mission (ESA \cite{esa97}). No variable type
could be assigned to this object and it was finally classified as unknown variable (flag 'U'). Aside from a determination of its V
magnitude, (B-V) and (U-B) color indices in a series of observations of stars with known accurate positions by Oja (\cite{oja85}),
there appears to be no further investigation of this star in the literature. HIP~7666 was included in the Grup d'Estudis
Astron\`{o}mics observing program (GEA \cite{gea99}) after the discovery of its eclipsing binary nature in a search for new variable stars 
in this field. Subsequent observations showed short-period oscillations 
superimposed on the eclipsing binary light curve, indicating that at least one of the components is a possible $\delta$ Scuti star. 
Since very few eclipsing binary stars with pulsating components of this type are known, we decided to perform follow-up observations to verify the nature of the observed light variations and to identify one or more pulsation periods.\\
Eclipsing binaries are a source of accurate information on the physical 
properties of the components such as stellar radii and masses. If one
of the components is in addition a (radial and/or non-radial) pulsating variable star, the system is of great scientific interest. The study of (non-radial) pulsators in such systems is very attractive 
for several more reasons:\\
   - the near equator-on view of the stellar surfaces is a favourable configuration
     for the detection of sectoral modes (i.e. non-radial modes with {\it l} = 
     {\it $\mid$m$\mid$});\\
   - during the eclipses, the components act as a spatial filter which causes
     phase and amplitude modulation of the non-radial mode and makes the mode
     identification more accurate (Mkrtichian et al. \cite{mkr02});\\
   - the pulsating component can be correctly identified if there is evidence 
     that the pulsations are most clearly present during one of both minima.
Rodr\'{\i}guez et al. (\cite{rod04}) recently summarized the status of the eclipsing
binaries with a $\delta$ Scuti component: only about a dozen cases are presently known.
About 30\% consist of eclipsing binaries with a classical $\delta$ Scuti pulsator while
the rest defines a distinct class of $\delta$ Scuti pulsators, called the oscillating EA
stars (Mkrtichian et al. \cite{mkr02}). 
The new data suggest that HIP~7666 is a normal main-sequence $\delta$ Scuti star in a detached eclipsing binary system. 
As such, the discovery of a $\delta$ Scuti pulsator in a 'clean' binary system not affected by mass accretion 
and/or transfer is extremely important. 

We present the newly collected photometric observations in section 2. In section 3, the photometric parameters are given and a Wilson-Devinney fit 
(hereafter WD, Wilson \cite{wils98}) is computed to obtain a synthetic eclipsing binary light curve that can be subtracted from the data to analyse the residuals. The short-period light oscillations detected in the residual data are discussed in section 4. In section 5 we summarize 
the results of this work.

\section{Observations and reduction}

HIP 7666 was observed in the V band from the Hostalets de Pierola 
Observatory\footnote{http://www.astrogea.org/jjs/index.html}\ using a 40cm Newtonian telescope for 21 nights,
and in the B and V bands from Monegrillo Observatory\footnote{http://www.astrogea.org/jvidal/index.html}\ also  
using a 40cm Newtonian telescope for 9 and 4 nights respectively. All our observations were acquired between 
August and November 2000. The complete  34-night data set spans a time interval of 85 days. The observational
log is listed in Table~\ref{Log}.
Both telescopes, having identical optical systems, were equipped with a SX Starlight CCD camera with a Sony ICX027BL chip cooled by a Peltier system to about -25\degr C. The field-of-view covered a sky region of 11.5'x7.7' and the pixel size was 1.80\arcsec {\rm x} 1.38\arcsec. Dark frames and flat fields were obtained, and the standard image processing  was carried out. The reduction was done using a software package called LAIA (Laboratory for Astronomical Image Analysis) and 
developed by Joan A. Cano\footnote{http://www.astrogea.org/soft/laia/laia.htm}\ . Since the field of HIP 7666 is not crowded, the technique of aperture 
photometry was applied to extract the differential magnitudes.

The brightness of HIP~7666 was measured with respect to HIP~7665 (=~HD~9976), while GSC 3671-1068 and GSC
3671-0834 served as check stars (Table~\ref{Stars}). Although the comparison star is actually an optical
pair (Dommanget \& Nys \cite{domm94}) the photometric measurements were not affected by the component. 
In fact, thanks to the small angular separation between both components, and the faintness of the secondary 
component ($\rho$=1.1\arcsec; $\Delta_{Hp}$ = 2.65 mag, ESA \cite{esa97}), HIP~7665 always appeared as a single
object on all the CCD frames. The overall rms scatter of the differences between the comparison star and the 
principal check star is 8 mmag. Because no suitable check star could be included in the frames - other
stars are at least 1 mag fainter than the comparison star - 8 mmag is an upper limit to the true photometric 
error. Since the check star is 2.5 times fainter than the comparison star, we can infer that the 
rms scatter of the differences between the variable and the comparison star is about 5 mmag.

\begin{table}
\caption{Observational log. In the fourth column HPO stands for Hostalets de Pierola Observatory and MO for Monegrillo Observatory}
\label{Log}
\begin{center}
\begin{tabular}{cccc}
\hline
Observation & Observation interval & Filter  &  Observatory \\
date & (HJD-2451000.0) & & code \\
\hline
8 Aug 2000 & 1765.553 - 1765.657 & V & HPO \\
9 Aug 2000 & 1766.363 - 1766.664 & V & HPO \\
9 Aug 2000 & 1766.376 - 1766.591 & V & MO \\    
10 Aug 2000 & 1767.423 - 1767.586 & V & HPO \\
10 Aug 2000 & 1767.364 - 1767.590 & V & MO \\    
12 Aug 2000 & 1769.526 - 1769.667 & V & HPO \\
13 Aug 2000 & 1770.407 - 1770.670 & V & HPO \\
14 Aug 2000 & 1771.387 - 1771.672 & V & HPO \\
15 Aug 2000 & 1772.364 - 1772.644 & V & HPO \\
16 Aug 2000 & 1773.384 - 1773.641 & V & HPO \\
17 Aug 2000 & 1774.351 - 1774.664 & V & HPO \\
18 Aug 2000 & 1775.397 - 1775.624 & B & MO \\
18 Aug 2000 & 1775.365 - 1775.660 & V & HPO \\
19 Aug 2000 & 1776.388 - 1776.604 & B & MO \\
19 Aug 2000 & 1776.409 - 1776.524 & V & HPO \\
20 Aug 2000 & 1777.395 - 1777.620 & V & HPO \\
24 Aug 2000 & 1781.375 - 1781.635 & V & HPO \\
25 Aug 2000 & 1782.376 - 1782.635 & V & HPO \\
28 Aug 2000 & 1785.350 - 1785.560 & B & MO \\
31 Aug 2000 & 1788.326 - 1788.598 & B & MO \\
1 Sep 2000 & 1789.334 - 1789.545 & B & MO \\
2 Sep 2000 & 1790.327 - 1790.586 & B & MO \\
4 Sep 2000 & 1792.323 - 1792.479 & B & MO \\
4 Sep 2000 & 1792.355 - 1792.582 & V & HPO \\
5 Sep 2000 & 1793.386 - 1793.522 & V & HPO \\
7 Sep 2000 & 1795.464 - 1795.464 & V & HPO \\
8 Sep 2000 & 1796.313 - 1796.587 & B & MO \\
8 Sep 2000 & 1796.372 - 1796.515 & V & HPO \\
9 Sep 2000 & 1797.361 - 1797.553 & V & HPO \\
10 Sep 2000 & 1798.370 - 1798.565 & B & MO \\
10 Sep 2000 & 1798.373 - 1798.490 & V & HPO \\
27 Oct 2000 & 1845.277 - 1845.496 & V & MO \\  
31 Oct 2000 & 1849.281 - 1849.439 & V & MO \\
31 Oct 2000 & 1849.373 - 1849.656 & V & HPO \\
\hline
\end{tabular}
\end{center}
\end{table}

\begin{table}
\caption{Catalogue data for the observed stars (ESA \cite{esa97})}
\label{Stars}
\begin{center}
\begin{tabular}{ccccc}
\hline
Star & Identifier & Sp. T. & $V_{T}$  &  $(B-V)_{T}$ \\
\hline
Var & HIP~7666 & A5 & 9.688 & 0.458\\
Comp & HIP~7665 & A2 & 8.997 & 0.103\\
Check1 & GSC~3671-1068 & G5 & 10.748 & 0.603 \\
Check2 & GSC~3671-0834 & -- & 11.442 & 0.553 \\
\hline
\end{tabular}
\end{center}
\end{table}

\section{The new binary system}

\begin{figure}
\resizebox{\hsize}{!}{\includegraphics{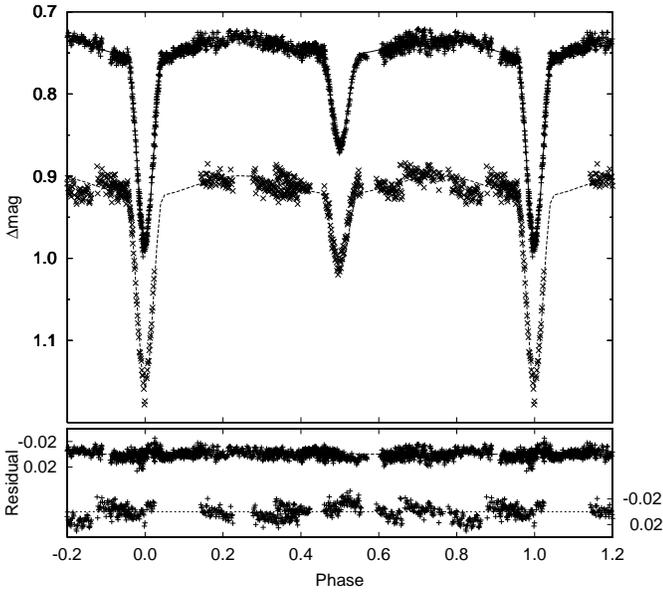}}
\caption{Upper panel: Light-curve of HIP 7666 folded with the $2.^{\rm d}37229$ period 
superimposed on the WD solution (upper curve: $\Delta V$, lower curve: $\Delta B$). 
Lower panel: Residuals after subtracting the WD fit from the original photometric data
(upper curve: $\Delta V$, lower curve: $\Delta B$).}
\label{Phase}
\end{figure}

Fig.~\ref{Phase} (upper panel) illustrates the light curves obtained in the V and B bands. The light curves
are typical of an eclipsing binary system of the detached or semi-detached type.
The observations reveal that HIP~7666 is an eclipsing binary system with a period of 2.37229 days 
which fades by 0.23 mag and 0.16 mag during, respectively, the primary and the secondary minimum. The total
(peak-to-peak) amplitude in V-light, including the ellipticity and reflection effects (causing the 
sinusoidal-like variations out-of-eclipse) in the close binary, 
is 0.25 mag. The measurements allowed us to compute a list of observed times of minima, presented in
Table~\ref{Minima}, from which we derived the following ephemeris by means of a linear least squares fit:

\begin{center}
${\rm Min.~I} = {\rm HJD~} 2451769.5615(11) + 2.^{\rm d}37229(8)\!\times\!{\rm E}$\\
\end{center}

\noindent{where E is the cycle number. Table~3 also lists the (O-C) residuals.}

\begin{table}
\caption{Observed heliocentric times of minima}
\label{Minima}
\begin{center}
\begin{tabular}{ccr}
\hline
HJD - 2450000 & Type & O-C \\
\hline
1769.5620 & I & 0.0005 \\
1775.4888 & II & -0.0034 \\
1775.4946 & II & 0.0025 \\
1781.4203 & I & -0.0026 \\
1782.6166 & II & 0.0075 \\
1788.5385 & I & -0.0013 \\
1845.4748 & I & -0.0001 \\
1845.4750 & I & 0.0002 \\
\hline
\end{tabular}
\end{center}
\end{table}

We made use of the WD code to model binary light curves (Wilson \cite{wils98}) to obtain a preliminary photometric solution for
the new binary. The available V- and B-curves were simultaneously modelled in the case of a detached 
system (mode 2). A semi-detached solution was also tested, but the best fits resulted with residuals an order of magnitude worse than for a detached configuration. We adopted a surface temperature of $8200~K$ (Schmidt-Kaler \cite{schmidt-kaler}) for the primary star corresponding to the star's global spectral type of A5 (Cannon \& Pickering \cite{HD1924}). Other assumptions were: a radiative bolometric albedo 
(A$_{1}$=1) and gravity darkening (g$_{1}$=1) for the primary, a convective bolometric albedo (A$_{2}$=0.5) and gravity 
darkening (g$_{2}$=0.32) for the secondary, and a logarithmic limb darkening law (Van Hamme \cite{vanham93}). We used the known 
spectral type to estimate the surface temperature because the wide-band photometry (Oja 1985) does not allow us to compute 
the reddening, which could be significant given its low Galactic latitude ($-9.7\degr$). Indeed, making use of the Str\"omgren 
measurements of the hotter star HIP~7729 (= HD~10054) (Westin \cite{westin}), which, with a parallax of $\approx$ 3 mas (ESA 
\cite{esa97}), is at about the same distance as HIP~7666 and in almost the same direction, we obtain a colour excess E(b-y) 
of 0.1 mag after application of the calibration procedure of the Str\"omgren photometric system (Moon \& Dworetsky \cite{moo85}). 
Because of the large uncertainty on the dereddened colour indices, we preferentially made use of the spectral type classification.

\begin{figure}
\resizebox{\hsize}{!}{\includegraphics{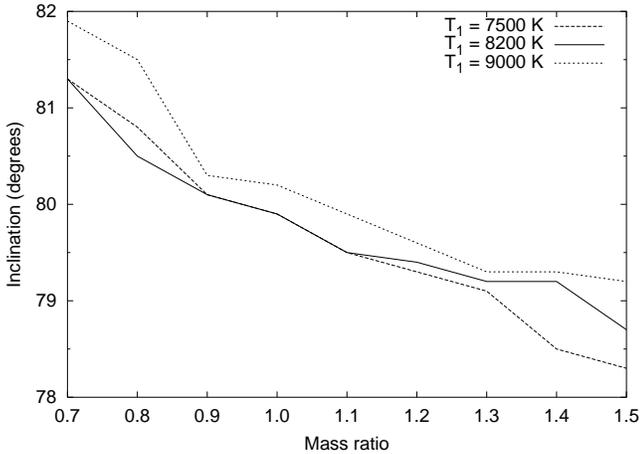}}
\caption{The inclination as a function of the mass ratio for three different values
of the primary's surface temperature.}
\label{Incl}
\end{figure}

Since no radial velocity information is available, we searched for the best solution by fixing the mass ratio 
$q = M_2/M_1$ and exploring a large number of values between 0.5 and 4.0. The surface temperature of the secondary, 
$T_{2}$, the gravitational potential for both components, the relative monochromatic luminosities $L_{1}$ and $L_{2}$, 
and the orbital inclination $i$ were considered to be free parameters. Since the eclipses are partial, and given the 
observational uncertainties, it was not possible to derive the photometric mass ratio in a reliable way and all $q$-values 
between 0.7 and 1.5 gave equivalent fits. However, the surface temperature of the secondary, $T_{2}$, does not depend 
much on the choice of $q$ within the given range. Table~\ref{Params} lists the parameters of the best fit when
$T_{1}$ equals 8200 K. We next investigated the parameter space using different values for the primary's surface temperature 
assuming an uncertainty of half a spectral class: adopting $T_{1}$ = $9000~K$ (A0-type star) or $T_{1}$ = $7500$ (F0-type 
star) however did not alter the quality of the fit significantly. In these cases, the secondary is $\approx$ 1500 K cooler
than the primary (see Table~\ref{Params}). In Fig.~\ref{Incl} the value of the inclination is represented as a function 
of the mass ratio $q$ for three values of the surface temperature $T_{1}$. Note that a mean value of $80 \pm 2\degr$ 
reflects well the range of possible values.

\begin{table}
\caption{Parameters of the binary models which gave equivalent best fits}
\label{Params}
\begin{center}
\begin{tabular}{cccc}
\hline
$T_{1}$ (K)& $T_{2}$ (K) & Sp. Type 2 & Range for $i$ \\
\hline
9000 & 7300 & A9-F0  & 82.0-79.0\\
8200 & 6700 & F3-F4  & 81.0-78.5\\
7500 & 6200 & F7-F9  & 81.0-78.5\\
\hline
\end{tabular}
\end{center}
\end{table}

Though modelling of the B- and V-curves did not allow us to constrain the physical parameters of the components uniquely or accurately, one can see that the match between one of these best solutions (we chose $q = 0.9$) and the observed light curves is very good (Fig.~\ref{Phase}, upper panel). This is especially obvious in the plot of the residuals where the peak-to-peak differences are of the order of 40 mmag at most (Fig.~\ref{Phase}, lower panel).

\section{Detection of short-period oscillations}

\begin{figure}
\resizebox{\hsize}{!}{\includegraphics{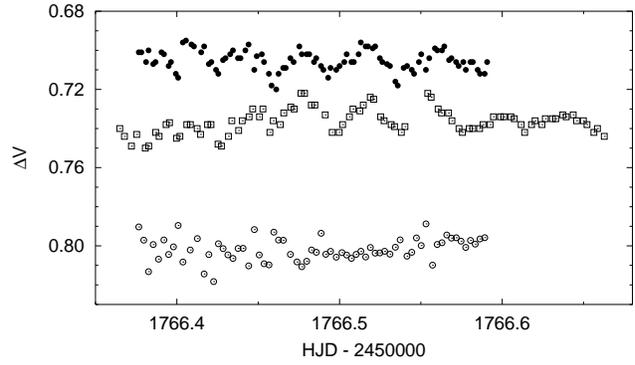}}
\caption{Comparison between the light curves of HIP~7666 taken from Monegrillo Observatory (filled dots, upper curve), 
and Hostalets de Pierola Observatory (open squares, middle curve). Both data series show the same light variations which 
are not present in the data of the comparison star with respect to the two check stars GSC 3671-1068 and GSC 3671-0834 collected at Monegrillo Observatory (open circles, bottom 
curve, note that these were shifted by a constant magnitude offset to fit in the plot). Note that since the check star is much fainter, the 
scatter is correspondingly higher.}
\label{HostMon}
\end{figure}

\begin{figure}
\resizebox{\hsize}{!}{\includegraphics{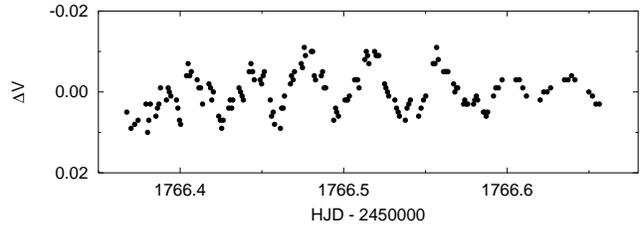}}
\caption{Light curve of HIP 7666 taken on JD 2451766.5 obtained 
after merging and averaging the data from Monegrillo and Hostalets de 
Pierola Observatories.}
\label{deltasc}
\end{figure}

In the plot of the residual light curves we notice periodic fluctuations with a maximum difference of 20 mmag. Despite 
their small amplitude, these are real as we could detect them during two different nights of simultaneous observations at Hostalets de Pierola Observatory and Monegrillo Observatory in Spain. Both observatories are about 300 km apart and are equipped with similar instrumentation (cf. Sect.~2). To rule out the possibility of observing variations caused by the comparison star, independent
photometry of HIP~7666 and HIP~7665 was performed with respect to the check stars GSC 3671-1068 and GSC 3671-0834 (cf.
Table~\ref{Stars}). Fig.~\ref{HostMon} illustrates the photometric runs taken from Hostalets de Pierola Observatory and Monegrillo
Observatory on the night corresponding to JD 2451766.5. The new photometric data confirmed that HIP~7666 exhibits
small-amplitude variations whereas the comparison star does not. Fig.~\ref{deltasc} shows the brightness variations for that
particular night, after merging and averaging the data from both observatories. It can be seen that the maximum amplitude variation
is about 20 mmag. The amplitude modulation present in this light-curve segment further suggests the presence of more than one frequency. The existence of short-period oscillations superimposed on the (eclipsing) binary light-curve is clearly demonstrated. Because of the short periodicity and the amplitude modulation, pulsations in one (or both) of the components are presumably the cause of these rapid light variations.

\subsection{The data during the phases of eclipse}

>From the V-light curve in Fig.~\ref{Phase} it seems that the residuals at the phase of primary minimum are somewhat 
larger than those of secondary minimum. No conclusion can be formulated from the incomplete B-light curve. Due to the mentioned above  uncertainties, however, we have no direct clue as to which component may be the pulsating star, but it may be supposed that the primary star which contributes most to the light is a good candidate. 
On the basis of the broad range in the spectral classification alone, both components may lie in the $\delta$ Scuti instability strip. The primary could be a main sequence pulsator, while the secondary (which is cooler) could be a more evolved $\delta$ Scuti star.

\subsection{The data out-of-eclipse}

\begin{figure}
\resizebox{\hsize}{!}{\includegraphics{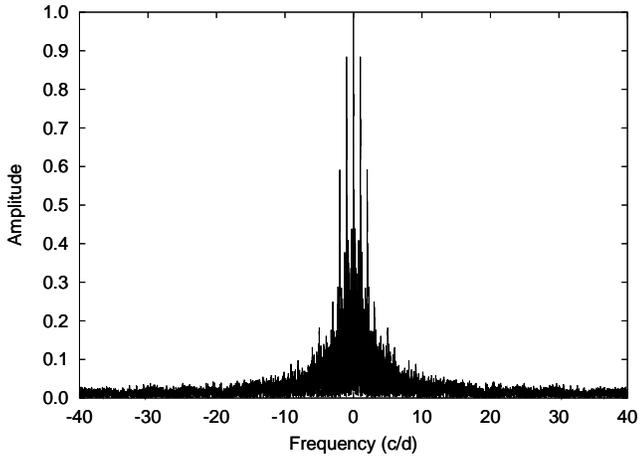}}
\caption{Spectral window for the observations of HIP~7666.}
\label{window}
\end{figure}

\begin{figure}
\resizebox{\hsize}{!}{\includegraphics{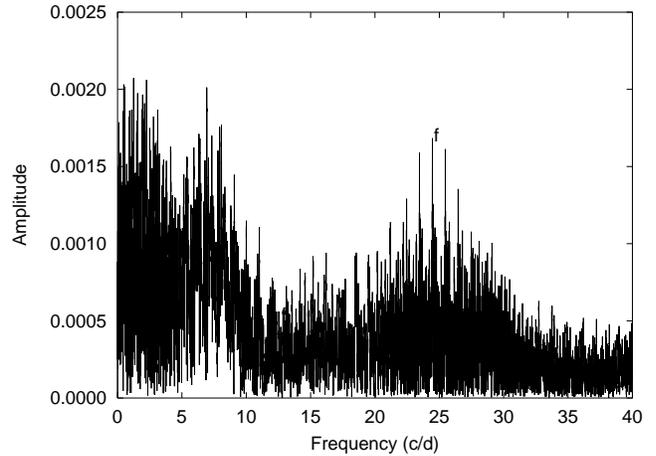}}
\caption{Power spectrum of the prewhitened out-of-eclipse data where the 24.46 c/d component is marked as {\bf f}.}
\label{spectrum}
\end{figure}

In order to perform a Fourier analysis in the frequency domain, we removed the WD fit from the original data set. First only the data out-of-eclipse were analysed using Period98 (Sperl~\cite{spe98}). The power spectrum is dominated by a 
lot of noise at frequencies smaller than 10 c/d, probably due to a combination of nightly zero-point shifts and other
observational errors. This situation did not change when the residual data during eclipse were included. In the less noisy 
part of the periodogram (in the range $> 10$ c/d) the highest peaks correspond to $24.463\pm 0.008$~c/d and its one day$^{-1}$ 
alias at $25.465\pm 0.008$~c/d with a semi-amplitude of $1.7$ mmag, introduced by the daily gaps in the observing pattern 
(see Fig.~\ref{window}). The quoted error bar corresponds to the half-width at half maximum of the peaks in the power 
spectrum. This is truly an upper limit of the error (Schwarzenberg-Czerny \cite{sch91}). The S/N ratio for this frequency is 4.1 and is just significant. This is illustrated by Fig.~\ref{spectrum}. 

\begin{figure}
\resizebox{\hsize}{!}{\includegraphics{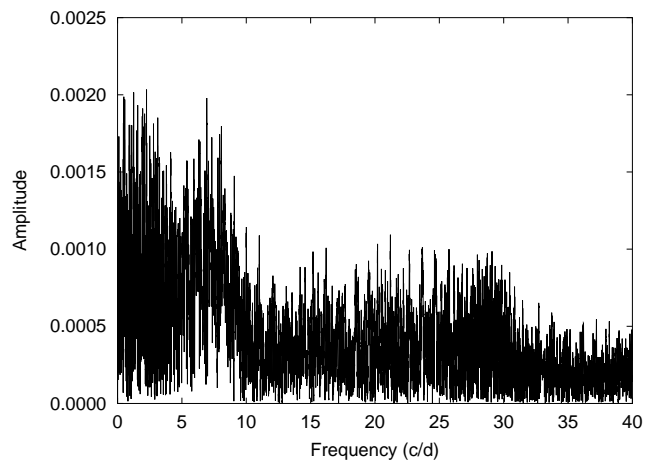}}
\caption{Power spectrum of the residuals data after prewhitening the $\delta$ Scuti pulsation signal.}
\label{residuals}
\end{figure}

\begin{figure}
\resizebox{\hsize}{!}{\includegraphics{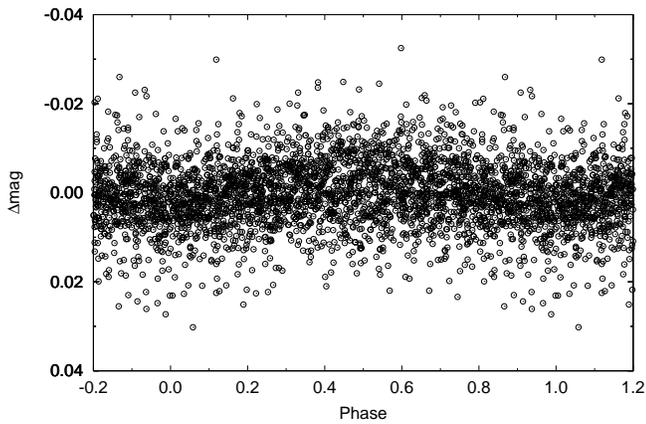}}
\caption{Folded light-curve of the prewhitened out-of-eclipse data folded on the 0.981 hour-period.}
\label{folded}
\end{figure}

Fig.~\ref{residuals} shows the power spectrum of the residual data after removing the $\delta$ Scuti pulsations, where the spectrum 
is noisy with most of the energy concentrated at frequencies below $10$ c/d. Fig.~\ref{folded} displays the folded 
light curve. Although the light curve of Fig.~\ref{deltasc} presents a vague hint of multiperiodicity, no other reliable
frequency could be found in that range of frequencies larger than $> 10$ c/d, as expected from the explicited observational error 
and the low S/N ratio. 

\section{Discussion}

Given the very short periodicity of about 1 hour and the location in the H-R diagram
(a star of spectral type A5V lies in the region where the classical instability strip
intersects the main-sequence), we conclude that at least one of the components of 
HIP~7666 is a pulsating star of type $\delta$ Scuti. Analysis of the light curve of
the eclipsing system further suggests that one component is lying in the hotter part
of the $\delta$ Scuti instability strip while the other component may just lie outside it, 
being probably cooler than its red border (cf. Breger \cite{bre00}). We therefore propose 
that the primary component is a $\delta$ Scuti pulsator.  

There are only a very few known normal $\delta$ Scuti stars that are members of detached eclipsing
binary systems (see Rodr\'{\i}guez~\cite{rod02};  Rodr\'{\i}guez et al.~\cite{rod04}). 
Recent investigations have demonstrated the existence of a group of A-F pulsating stars located 
in the $\delta$ Scuti instability strip and with the same observational characteristics as
$\delta$ Scuti stars but in a different evolutionary stage. This new group consists of 
mass-accreting main-sequence pulsating components of semi-detached Algol-type binaries
(Mkrtichian et al. \cite{mkr02}). Presently thirteen members have been discovered, including the latest discoveries (Caton \cite{cat04}, Kim et al. \cite{kim04}, \cite{kim05}, Lampens et al. \cite{lam04}), but excluding WX~Eri (Arentoft et al. \cite{are04}). They have orbital periods ranging from 1 to 7 days and pulsation periods from 22 min to 6.5 hrs. One complex example is RZ Cas (Mkrtichian et al.~\cite{mkr03}). 

Our observations suggest that HIP~7666 is a classical $\delta$ Scuti star in a detached eclipsing binary system, which probably does not belong to the previous class. Nevertheless, the interest of such a discovery is high since these objects allow one to study the undisturbed pulsation characteristics while also providing the basic stellar properties needed to better constrain the pulsation models. Further observations are necessary, in particular 
high-resolution spectroscopic data would be helpful to confirm the precise nature of the binary system and to fully explore the nature of the detected pulsations. We would welcome new collaborations in this respect, as we have already started an extensive, multi-site photometric campaign.

\begin{acknowledgements}
The authors are pleased to acknowledge Prof. R. Wilson (University of Florida) for the Wilson-Devinney code. P. Lampens acknowledges support from the Fund for Scientific Research (FWO) - Flanders (Belgium) (project G.0178.02). We further thank Joan A. Cano and Rafael Barber\'a for writing the software for obtaining and reducing the CCD frames. We are also greateful to the anonymous referee whose comments helped to improve this paper. We made use of the Simbad database, operated by the {\it Centre de Donn\'ees astronomiques de Strasbourg} (France), and the ADS service.
\end{acknowledgements}

\end{document}